\documentclass[aps,prb,twocolumn,superscriptaddress,showpacs]{revtex4-1}
\usepackage{mathbbol,amsmath,amsfonts,amssymb,bm,color,dsfont}
\usepackage{graphicx,psfrag,array}

\def\klambda{{{\ket{\lambda}}}}
\def\klambdap{{{\ket{\lambda'}}}}
\def\ek{\mathfrak{e}}
\def\psiM{\ket{\psi_{1/M}}}
\def\tpsiM{\ket{\tilde\psi_{1/M}}}
\def\mS{\mathfrak{S}}
\def\cN{{\cal N}}
\def\cO{{\cal O}}
\providecommand{\ignore}[1]{}
\providecommand{\aucmnt}[1]{#1}
\def\be{\begin{equation}}
\def\ee{\end{equation}}
\renewcommand{\aucmnt}[1]{}

\newcommand{\ket}[1]{| #1 \rangle}

\newcommand{\Comment}[1]{}

\newcommand{\Eq}[1]{Eq.~(\ref{#1})}

\makeatletter
\let\save@mathaccent\mathaccent
\newcommand*\if@single[3]{%
  \setbox0\hbox{${\mathaccent"0362{#1}}^H$}%
  \setbox2\hbox{${\mathaccent"0362{\kern0pt#1}}^H$}%
  \ifdim\ht0=\ht2 #3\else #2\fi
  }
\newcommand*\rel@kern[1]{\kern#1\dimexpr\macc@kerna}
\newcommand*\widebar[1]{\@ifnextchar^{{\wide@bar{#1}{0}}}{\wide@bar{#1}{1}}}
\newcommand*\wide@bar[2]{\if@single{#1}{\wide@bar@{#1}{#2}{1}}{\wide@bar@{#1}{#2}{2}}}
\newcommand*\wide@bar@[3]{%
  \begingroup
  \def\mathaccent##1##2{%
    \let\mathaccent\save@mathaccent
    \if#32 \let\macc@nucleus\first@char \fi
    \setbox\z@\hbox{$\macc@style{\macc@nucleus}_{}$}%
    \setbox\tw@\hbox{$\macc@style{\macc@nucleus}{}_{}$}%
    \dimen@\wd\tw@
    \advance\dimen@-\wd\z@
    \divide\dimen@ 3
    \@tempdima\wd\tw@
    \advance\@tempdima-\scriptspace
    \divide\@tempdima 10
    \advance\dimen@-\@tempdima
    \ifdim\dimen@>\z@ \dimen@0pt\fi
    \rel@kern{0.6}\kern-\dimen@
    \if#31
      \overline{\rel@kern{-0.6}\kern\dimen@\macc@nucleus\rel@kern{0.4}\kern\dimen@}%
      \advance\dimen@0.4\dimexpr\macc@kerna
      \let\final@kern#2%
      \ifdim\dimen@<\z@ \let\final@kern1\fi
      \if\final@kern1 \kern-\dimen@\fi
    \else
      \overline{\rel@kern{-0.6}\kern\dimen@#1}%
    \fi
  }%
  \macc@depth\@ne
  \let\math@bgroup\@empty \let\math@egroup\macc@set@skewchar
  \mathsurround\z@ \frozen@everymath{\mathgroup\macc@group\relax}%
  \macc@set@skewchar\relax
  \let\mathaccentV\macc@nested@a
  \if#31
    \macc@nested@a\relax111{#1}%
  \else
    \def\gobble@till@marker##1\endmarker{}%
    \futurelet\first@char\gobble@till@marker#1\endmarker
    \ifcat\noexpand\first@char A\else
      \def\first@char{}%
    \fi
    \macc@nested@a\relax111{\first@char}%
  \fi
  \endgroup
}
\makeatother

\begin{document}

\title{Zero modes, Bosonization and Topological Quantum Order: \\ The Laughlin State in Second Quantization}

\author{Tahereh Mazaheri}
\affiliation{Department of Physics, Washington University, St.
Louis, MO 63160, USA}
\author{Gerardo Ortiz}
\affiliation{{Department of Physics, Indiana University, Bloomington,
IN 47405, USA}}
\author{Zohar Nussinov}
\affiliation{Department of Physics, Washington University, St.
Louis, MO 63160, USA}
\author{Alexander Seidel}
\affiliation{Department of Physics, Washington University, St.
Louis, MO 63160, USA}

\date{\today}
\begin{abstract}
We introduce a ``second-quantized'' representation of the ring of 
symmetric functions to further develop a purely second-quantized -- or ``lattice'' --
approach to the study of zero modes of frustration free
Haldane-pseudo-potential-type Hamiltonians, which in particular stabilize Laughlin ground states. We present three applications of this formalism.
We start demonstrating how to systematically construct all
zero-modes of Laughlin-type parent Hamiltonians in a framework
that is free of first-quantized polynomial wave functions, and show that
they are in one-to-one correspondence with dominance patterns.
The starting point here is the pseudo-potential Hamiltonian in ``lattice form'',
stripped of all information about the analytic structure of Landau levels (dynamical momenta).
Secondly, as a by-product, we make contact with the bosonization method,
and obtain an alternative proof for the equivalence between bosonic
and fermionic Fock spaces.
Finally, we explicitly derive the second-quantized version of Read's non-local (string) 
order parameter for the Laughlin state, 
extending an earlier description by Stone. Commutation relations between the
local quasi-hole operator and the local electron operator are generalized to
various geometries.
\end{abstract}
\maketitle

\section{Introduction}

The physics of electrons in a strong, external,  magnetic field
harbors a great multitude of interesting phases of strongly
correlated electrons. For a subset of these phases, 
representative wave functions can be given that 
have sufficiently simple analytic properties such that a parent
Hamiltonian can be constructed. This includes\cite{haldane_hierarchy, TK} the experimentally
relevant Laughlin\cite{Laughlin} states, as well as the non-Abelian Moore-Read state,\cite{MR}
which may explain the plateau at filling factor $\nu=5/2$,\cite{fivehalves} as well as the entire 
Read-Rezayi series,\cite{RR} the Gaffnian\cite{gaffnian} state which is presumed critical,  as well as a variety of
multi-component wave functions.\cite{halperin_helv, HR}
Certainly, the dichotomy between analytic wave functions and their parent Hamiltonians 
is a key component of the theory of fractional quantum Hall states.
At the same time, for a great wealth of quantum Hall states that are understood through 
hierarchical,\cite{haldane_hierarchy, halperin_hierarchy} composite fermion,\cite{jain} or 
complementary field theoretical constructions,\cite{wen_hier_edge} we do not enjoy the luxury 
of sufficiently ``special'' microscopic wave functions
that can be seen to be exact eigenstates of suitable local Hamiltonians.

In the more fortunate cases, it is the existence of special analytic ``clustering'' 
properties\cite{haldane_hierarchy, greiter, RR, bernevig1, bernevig2, WW1, WW2} of first 
quantized wave functions that makes construction of a parent Hamiltonian possible. 
Interestingly, it has recently been suggested by Haldane that the well-known analytic 
properties of such wave functions may be deceptive, in the sense that they become 
meaningless after Landau level projection.\cite{Haldane11}
Indeed, as a result of such projection, one discards all the degrees of freedom of the problem
relating to dynamical momenta (which determine the structure of Landau levels) and 
keeps only the degrees of freedom associated with guiding centers. The full  Hilbert space is isomorphic
to the tensor product
\be\label{HS}
{\cal H}_\pi\otimes {\cal H}_\omega,
 \ee
  with factors describing degrees of freedom belonging to dynamical momenta ($\pi$) and 
  guiding centers ($\omega$), respectively. In principle, the problem of working out the 
  eigenstates of a (projected) interacting Hamiltonian could be naturally reduced to the 
  second factor, were it not for the fortuitous circumstance that  looking at the problem 
  in the full Hilbert space with the ``right'' Landau level structure,
  ground state wave functions are sometimes seen to have exceptional analytic properties.
Often, however, this may not be the case. In view of the above, it may not be surprising
that for many quantum Hall states, some as basic as the Jain $\nu=2/5$ state, no representative
wave functions are known with analytic properties that are ``special'' enough to allow the construction 
of a suitable parent Hamiltonian. Moreover, ``quantum Hall like'' Hamiltonians have 
become fashionable in contexts where the traditional Landau level structure, which 
provides the basis for these analytic properties, is absent, and is replaced\cite{qi} by a
 basis of Wannier states in the flat ``Chern band'' (having non-zero Chern number) of a ``fractional Chern insulator''.
There has been much interest recently in suitable flat band systems.
\cite{Tang11,Sun11,Neu11,Sheng11,qi, Wang11,Regnault11,Ran11,Bernevig12,WangPRL12,WangPRB12,Wu12,Laeuchli12,Bergholtz12,Sarma12,Liu13,chen13, scaffidi,wang2014}

In such a context, the traditional analytic wave functions are meaningless, if still useful for 
the construction of solvable models. Instead, only a manifestly Landau level projected, 
or ``guiding center'' representation of quantum many-body states is meaningful.
A formalism that naturally implements Landau level (or flat band) projection is obtained 
by passing to second quantization. For example, for the Haldane 
pseudo-potentials\cite{haldane_hierarchy}
one obtains second quantized expressions of the form (see, e.g., Ref. \onlinecite{ortiz} for the two-body case in any geometry, or,  Ref. \onlinecite{CHLee}
for general $n$-body generalizations on the cylinder ):
\be\label{pseudo}
\begin{split}
  H_m &= \sum_{R} {T^m_R}^\dagger T^m_R\,,\\
  T_R^m&=\frac{1}{2} \sum_{x} \eta^m_{R,x} \, c_{R-x} c_{R+x}\,,
\end{split}
\ee
 where $m$ refers to the $m$th Haldane pseudo-potential, 
 $R$ and $x$ run over integer and half-integer values, with $x$ constrained via 
 $(-1)^{2x}=(-1)^{2R}$, $c_r$ annihilates a 
 boson (fermion) in the $r$th Landau level orbital for $m$ even (odd),
 and the $\eta_{R,x}^m$ are form factors that depend on geometry,
 and are polynomial in $x$ for the sphere, disk, or cylinder geometry.\cite{ortiz} 
 In the present paper, we will always work in these geometries. Notice that 
 $H_m$ is a separable-potential Hamiltonian. \cite{ortiz}

 It is well-known\cite{haldane_hierarchy} that the $\nu=1/M$ Laughlin state
 is the unique zero energy (zero mode) ground state at filling factor $1/M$ of the Hamiltonian
 \be\label{parent}
 H^M=\sum_{\substack{0\leq m<M\\ (-1)^m=(-1)^M}} V_m H_m\,,
 \ee
 with positive real numbers $V_m$.
 This and other properties of the Laughlin state are well established
 using the analytic (polynomial) structure of first quantized wave-functions,
 thus, making use of the embedding of the Landau level(s) into the larger
 Hilbert space \eqref{HS},  which endows lowest Landau level wave functions 
 with their polynomial character.
 
 In view of the above, one may, however, wonder if such embedding is {\em necessary}
 to understand the (zero mode) properties of the Hamiltonian \eqref{pseudo}, \eqref{parent}
 when given in the above, manifestly projected, second-quantized form.
 Understanding the problem without the introduction of spurious (as far as interactions
 are concerned) degrees of freedom may not only be pleasing from a mathematical point of 
 view, but the abandonment of a manifestly polynomial wave-function structure requires the 
 development of a new route of attack that may well be beneficial in the broader context of 
 constructing quantum Hall parent Hamiltonians. Moreover, one observes that written in the 
 form \eqref{pseudo}, the Hamiltonian belongs to a particular breed of frustration-free lattice 
 Hamiltonians, which have generally attracted much interest in recent 
 years.\cite{AKLT, nachtergaele, Kennedy92, Bravyi09, NdB, Yoshida, CJKWZ, MZ, schuch, Bravyi, Darmawan}   
 We call this class of generalized pseudo-potential systems, {\it Loop-algebraic Hamiltonians} 
 because of the underlying algebra the operators $T_R^m$ satisfy. \cite{ortiz}
 
 The connection with frustration-free lattice Hamiltonians may not be surprising, in view of the 
 recently discovered matrix-product structure of the Laughlin state.\cite{dubail, zaletel}
 Unlike in other well-known examples of frustration-free models, the $H_m$'s in \eqref{pseudo}
 are not strictly finite ranged in the lattice basis. It would thus seem that studying the existence and 
 properties of zero modes of general Hamiltonians of the form  \eqref{pseudo}, \eqref{parent}, 
 with generic, not necessarily short-ranged coefficients $\eta^m_{R,x}$, is a much harder problem 
 than for ordinary, finite-range frustration-free models.
 Nonetheless, some general mathematical statements have been obtained in Ref. \onlinecite{ortiz}.
 Under quite general circumstances, which we shall not repeat here but which apply for the 
 $\eta^m_{R,x}$ corresponding to Haldane pseudo-potentials, any zero mode of the Hmiltonian
 \eqref{parent} can be ``inward squeezed'' from one or several partitions satisfying a 
 ``generalized Pauli-principle''.\cite{bernevig1, bernevig2} We review details below. 
 This statement generalizes a fact that is known for many quantum Hall wave functions, 
 especially, those with polynomial wave functions characterized by certain clustering 
 properties,\cite{bernevig1,bernevig2}
 to zero modes of essentially {\em any} Hamiltonian  of the form  \eqref{pseudo}, \eqref{parent}
 (with natural generalizations to higher-body terms). This includes the lattice Hamiltonians 
 constructed in Refs. \onlinecite{Nakamura1, Nakamura2}, whose ground states are not described 
 by polynomials with nice clustering properties.
 For the Hamiltonian \eqref{pseudo}, \eqref{parent}, one then immediately obtains\cite{ortiz} that there 
 are no zero modes at filling factors $\nu>1/M$, and that there is at most one zero mode at filling factor
 $\nu=1/M$ (assuming the topology of the sphere, cylinder, or disk). A zero mode at
 $\nu=1/M$ would thus always have an ``incompressible character'' for the given class of 
 Hamiltonians, in the sense that at finite system size, a state at $\nu>1/M$ necessarily has 
 positive energy (although we make no statement here about the thermodynamic limit). For the 
 purpose of this paper, it is beneficial to use the term ``incompressible'' in the above sense.
 
Specializing to the case of Haldane peudo-potentials and taking, for now, the existence of an 
incompressible  zero mode at $\nu=1/M$ as given
 (which then of course is just the $1/M$-Laughlin state), second quantized operators have 
 been identified\cite{ortiz} that generate new zero modes, at lower filling factor $\nu < 1/M$, when acting 
 on any given zero mode.
 These operators depend on a positive integer $d$, and with normalization conventions that 
 belong to the infinitely ``thick'' cylinder geometry and that we will review below, can be written as
 \be\label{Od}
   \cO_{d} = \sum _{r}  c_{r+d}^{\dagger}c_{ r}^{\;}\, .
\ee
The operators ${\cal O}_d$ generate a commutative unital algebra
 $\cal A$. 
 Once it is known that: 1) there is a special ``incompressible'' zero mode $\ket{\psi_{1/M}}$ at filling factor
 $1/M$, and 2) {\em every} zero mode can be written in the form $\hat a\ket{\psi_{1/M}}$ with $\hat a\in {\cal A}$
 (i.e., $\hat a$ a linear combination of products of ${\cal O}_d$'s), then we can say that every property
 of zero modes known from the first quantized analytic wave function approach can also be 
 understood from a second-quantized or ``lattice'' point of view, which is manifestly ``guiding-center''
 and for which the starting point is given by the operators $T^m_R$ of \Eq{pseudo}.
 The first statement, the existence of a state at  filling factor $1/M$ that is annihilated by all
 operators $T^m_R$ with $m<M$ and $m$ even (odd) for $M$ even (odd), is demonstrated
 in a purely second quantized approach by work in parallel, Ref. \onlinecite{Chen14}. 
 This is done by giving a recursive definition for the Laughlin state in the lattice basis, and 
 demonstrating that it has the zero mode property, making no reference to first quantized 
 polynomial wave functions. Here we will prove the second statement in a similarly wave-function 
 free approach, thus completing the program of describing zero modes of Haldane 
 pseudo-potentials and related Loop-algebraic Hamiltonians (stabilizing the Laughlin state) in an intrinsically 
 second-quantized language.
 
 We point out that application of the operator ${\cal O}_d$ corresponds\cite{ortiz} to the 
 multiplication of certain symmetric polynomials (power-sum symmetric polynomials) in 
 first quantization, though we will not make direct use of this fact. The zero mode excitations 
 created in this way may be viewed as edge excitations, although they may of course live 
 arbitrarily deep in the bulk of the system (in particular, for large $d$, and are then better 
 thought of as quasi-hole excitations).  Indeed, there is an obvious connection between the form of the operators
 ${\cal O}_d$, and the boson operators in the traditional bosonization scheme (even though 
 the operators $c_r$ may themselves be bosonic or fermonic here, depending on the situation). 
 It is thus worth emphasizing that all our results apply to finite systems of arbitrary size, 
 and require no long wavelength (effective field theory) limit to be taken.

The remainder of the paper is organized as follows. 
We derive the completeness of the zero modes generated by the operators
$\cO_d$ in Secs. \ref{prelim} and \ref{proofs},
 We then proceed to make contact with the standard bosonization scheme in Sec. \ref{boson}, by 
 observing that as a byproduct of our results,  one obtains an alternative proof of  a standard 
 theorem of the bosonization method, which states the equivalence of the fermonic and 
 the bosonic Hilbert space. 
 
Finally, in Sec. \ref{OP}, we give yet another application of our second-quantization  
formalism, which is to express the
local quasi-hole operator explicitly in terms of the second quantized electron operator.
The quasi-hole operator is the central ingredient to Read's non-local (string) order parameter of  
Laughlin quantum Hall states.\cite{readOP} We extend the commutation relations between this operator 
and the local electron
operator given by Stone\cite{stone_book} for the disk geometry to the cylinder and the sphere. 
 
 At the technical level,  a main contribution of the present work is the 
 representation of the ring of universal symmetric functions through an
 embedding into the algebra of canonical commutation- or anti-commutation-relations.
 This embedding plays a crucial role in all of our results, and maps alternatively
 power-sum or elementary symmetric polynomials to fairly simple expressions
 in terms of fermionic or bosonic ladder operators, obeying well-known
 non-trivial relations of the Newton-Girard type.
  We believe that this mapping
 could be of quite general value beyond the applications given here.

 
\section{Definitions and preliminary considerations\label{prelim}}

\subsection{Statement of the problem}

We will, in the following, work with a Hilbert space defined by a half-infinite lattice
of orbitals $\phi_r$ labeled by an integer $r\geq 0$. This may describe the lowest
Landau level of a system with either disk geometry, or a half-infinite cylinder,
or, if an additional upper cutoff is introduced, a sphere.
Zero-modes of the Laughlin state parent Hamiltonians in any two of these geometries
are in one-to-one correspondence (up to the aforementioned cutoff, if one of the geometries is 
the sphere). In second quantization, these zero modes differ only by normalization 
factors assigned to each basis state,\cite{ortiz} as we review in Sec. \ref{OP}.
Here, we will give all operators using the normalization conventions of the infinite-radius 
cylinder, which are simplest. 
Fixing a certain value for $M$, zero modes can then be 
characterized as states annihilated
by all operators\cite{ortiz}
\be\label{QR}
  Q^m_R=\sum_{\substack {x\\ (-1)^{2x}=(-1)^{2R}}}  x^m \, c_{R-x} c_{R+x}\,,
\ee
where the $c_r^{\;}$, $c_r^\dagger$ are bosonic (fermionic) ladder operators when $M$ is 
even (odd),
$m$ runs over even (odd) values satisfying $0\leq m<M$, $R$ runs over non-negative 
half-integer values, $x$ runs over all values such that $R\pm x$ is integer, and we 
use the convention $c_r^{\;}=c_r^\dagger=0$ for $r<0$. The operators $Q^m_R$ may 
be thought of as linear combinations
of the $T^m_R$ in \Eq{pseudo} in the limit of an infinitely thick cylinder. These linear 
combinations
are taken for convenience to yield the simple monomial factors of \Eq{QR}. In general, the 
zero mode condition can be stated in terms of the original operators $T^m_R$ or any linearly
independent combination thereof.
We will thus be interested in finding {\em all} states $\ket{\psi}$ in the Hilbert space that are characterized by
the following algebraic property:
\be
\begin{split}\label{zeromode}
      Q_R^m \ket{\psi}=0\quad &\forall \ R=0,\frac 12, \dotsc,\\
      & \forall  \ 0\leq m<M, (-1)^m=(-1)^M\,,
\end{split}
\ee
which we will refer to as the {\em zero mode property} (for fixed $M$). These states $\ket{\psi}$ 
constitute the low-energy subspace of the Hilbert space ${\cal H}_\omega$.
We define the filling factor of any zero mode $\ket{\psi}$ of $N$ particles as
\be
     \nu=\frac{N-1}{r_{\sf max}}\,,
\ee
where $r_{\sf max}$ is the highest orbital index among the orbitals occupied in the
zero mode, i.e. $r_{\sf max}=\max \,\{r| \langle\psi |c^\dagger_r c_r^{\;}|\psi\rangle\neq 0\}$.

For a much more general class of problems defined through general deformations
of the operators \eqref{QR},
it is known\cite{ortiz} that zero modes can exist only for filling factor $\nu\leq 1/M$.
It is further known, for the same general class of problems studied in Ref. \onlinecite{ortiz},
that {\em if} there exists a zero mode at filling factor $1/M$, it is unique, and is of the form
\be \label{dominance}
   \ket{\psi_{1/M}}= \ket{\tilde \psi_{1/M}}+\sum_{\lambda}  \! ' \  C_\lambda \ket{\lambda},
\ee
where $\sum_{\lambda}  \! ' $ excludes the term $\lambda=\tilde \psi_{1/M}$, and 
\be
 \ket{\tilde \psi_{1/M}}= \ket{1\underbrace{0\dots 0}_{M-1}1\underbrace{0\dots 0}_{M-1}10\dotsc}
\ee
is the  ``thin cylinder/torus pattern''\cite{RH,  BK1, seidel05, SL,seidellee07, BK2, BK, SY08, ABKW, BHHKV, Seidel10, LBSH, SY11, papic, WS} or ``root partition'',\cite{bernevig1, bernevig2, BH3,  regnault, ortiz} where 
$1$s and $0$s denote occupation numbers  of the orbitals created by $c_r^\dagger$,
with $1$s separated by $M-1$ zeros.  
The $\ket{\lambda}$'s denote other occupation number eigenstates. 
As the occupation number eigenstates form a basis of the Hilbert space, the statement 
of \Eq{dominance} becomes non-trivial only with the additional information that 
$C_\lambda\neq 0$ only for occupation number configurations $\lambda$ that are dominated,
in the usual sense,\cite{RH} to be reviewed below, by the configuration $\ket{\tilde \psi_{1/M}}$.

For the problem at hand, with $Q_R^m$ defined as given in \Eq{QR},
it is further known that a unique zero mode exists at filling factor $1/M$.
This is simply the $1/M$-Laughlin state $\ket{\psi_{1/M}}$.
We note that the existence of such a state can, if desired, be derived solely from the zero mode
condition \eqref{zeromode}, and algebraic properties of the operators $Q_R^m$, i.e., in the algebraic,
wave-function-free language preferred here (See work in parallel\cite{Chen14}).
Our goal here will be to prove the following statement:

{\em
{\bf Theorem 1}. Every zero mode is the linear combination of states given by products of operators
${\cal O}_d$ ($d>0$) acting on the special zero mode $\ket{\psi_{1/M}}$.
}

In slightly more technical terms, consider the algebra $\cal A$ generated by the (commuting) 
operators ${\cal O}_d$, $d>0$. Then we have that the zero mode subspace 
(of ${\cal H}_\omega$) $Z$ is obtained
by applying all elements $\hat a$ of $\cal A$ to the highest filling factor zero mode $\ket{\psi_{1/M}}$:
\be
Z={\cal A} \ket{\psi_{1/M}}=\{{\hat a}  \ket{\psi_{1/M}} \ , \mbox{ with } \ \hat a \in {\cal A}\}\,.
\ee

\subsection{A second-quantized representation for the ring of symmetric polynomials}

While our goal is here to establish techniques to proof Theorem 1 in an intrinsically 
second-quantized fashion, it would be remiss if we did not make contact with 
the usual first quantized procedure every now and then, for reasons of transparency and 
pedagogy.  The key property of the operators ${\cal O}_d$
is that they produce new zero modes at lower filling factor and same particle number when acting
on given zero modes. This is a simple consequence of the commutation relations\cite{ortiz}
between the ${\cal  O}_d$ and the $Q_R^m$, or the original $T^m_R$, \Eq{pseudo}.
However, it also follows from the fact that the action of the operator ${\cal O}_d$ on  a given state, 
in first quantized language,
is to multiply the wave function with ``power-sum" symmetric polynomials,
\be\label{powersum}
p_d = \sum_{i=1}^{N} z_i^d\,.
\ee
The identification of zero modes with symmetric polynomials has a long 
tradition.\cite{haldane_prange, wen_edge, RR96}
As emphasized, here we wish to bypass this language entirely.
Instead we seek a way to establish a one-to-one correspondence between zero modes
and dominance patterns satisfying certain 
rules\cite{RH, seidel05, BK2, bernevig1,bernevig2,regnault} that follows directly
from the algebraic zero mode definition, \Eq{zeromode}. It turns out, however, that a direct
proof of Theorem 1 as stated above is quite cumbersome.
Our strategy is to restate the problem in terms of a different set of zero modes generating operators,
namely those that correspond, in first quantization, to the multiplication of 
wave functions with {\em elementary} symmetric polynomials:
\be\label{elementary1}
{s_d= }\sum_{1\leq i_1 < \dotsc <i_d\leq N} z_{i_1}z_{i_2}\dotsm z_{i_d}\,.
\ee
We postulate, and show later,  that this is facilitated by the following operators:
\be\label{elementary2}
      e_{d} = \frac{1}{d!} \sum _{ r_{1} \dotsc r_{d}}  c_{r_{1}+1}^{\dagger}c_{r_{2}+1}^{\dagger}\dotsm 
      c_{r_{d}+1}^{\dagger}c_{ r_{d}}^{\;}\dotsm c_{r_{2}}^{\;}c_{r_{1}}^{\;}\,.
\ee

For pedagogical purposes and to develop intuition, let us illustrate the 
action of the operators ${\cal O}_d$ and $e_d$ when $d=2$ on a simple state, such as 
the root partition $\ket{\tilde{\psi}_{1/3}}=\ket{1001001001000\dotso}$, with $N=4$ and $r_{\sf max}=9$
\begin{eqnarray}\label{O2}
{\cal O}_2 \ket{\tilde{\psi}_{1/3}}=&\ket{0011001001000\dotso} +\ket{1000011001000\dotso} \nonumber \\  
&+ \ket{1001000011000\dotso}+\ket{1001001000010\dotso},\nonumber\\ 
\end{eqnarray} 
\begin{eqnarray}\label{e2}
{e}_2 \ket{\tilde{\psi}_{1/3}}&=\frac{1}{2!}&(\ket{0100101001000\dotso}+\ket{0101000101000\dotso} \nonumber \\ 
&+& \ket{1000100101000\dotso}+\ket{0101001000100\dotso}\nonumber\\
&+&    \ket{1000101000100\dotso}+\ket{1001000100100\dotso}).\nonumber\\
\end{eqnarray} 

In this work, we wish to avoid making direct contact with the polynomial language.
Rather, we want make sure that the logic we follow, and generality of our results,  
are entirely independent of the 
analytic polynomial wave functions. Therefore, we will {\em not} proceed by showing
any connection between  Eqs. \eqref{elementary1} and \eqref{elementary2}, though it would
not be difficult to do that.
Rather, we seek to directly establish the algebraic relations between the operators
\eqref{Od} and \eqref{elementary2}.
These must be the same as those between the associated symmetric polynomials, if the 
aforementioned associations are correct. These are the Newton-Girard {\it operator} relations:
($e_0=0$ and $e_1={\cal O}_1$)
\be\label{NG}
    d\,e_d+ \sum_{k=1}^{d}(-1)^{k}\cO_ke_{d-k} = 0\,.
\ee
That the $\cO_d$ and $e_d$ are indeed so related is shown in Appendix \ref{appA}.
Consider again the simple example above with $d=2$. The Newton-Girard operator 
relations imply that $2e_2={\cal O}_1^2-\cO_2$, and 
it is easy to see that the action of $\cO_1^2$ on $\ket{\tilde\psi_{1/3}}$ 
does indeed produce every term in \Eq{O2}, as well as {\em twice} every term
in \Eq{e2},
indeed satisfying the required relations.

The relations \eqref{NG} show in particular that,  by induction, every $e_d$ is expressible in terms
of ${\cal O}_d$'s, and vice versa. From this fact, we in particular infer that the $e_d$'s 
must have the following in common with the ${\cal O}_d$'s (for which these statements 
are already known\cite{ortiz}):
\begin{itemize}
\item
The $e_d$'s all commute.

\item
When acting on zero modes, the $e_d$'s generate new zero modes.

\item 
The $e_d$'s generate the same algebra $\cal A$ as the ${\cal O}_d$'s.

\end{itemize}

From the last statement, we see that in Theorem 1, we may replace
${\cal O}_d$ with $e_d$ without changing the meaning. It is in this form that we will
prove Theorem 1.

It is a remarkable property of the Newton-Girard relations that they do not depend 
on the number of variables $N$, even though the polynomials ${\cal O}_d$ and $s_d$ 
do (where the latter vanish for $d>N$). Related to that, one can define the ring of 
``universal symmetric functions'',\cite{macdonald} which may be thought of as the limit 
$N\rightarrow \infty$ of the ring of symmetric polynomials for fixed $N$.
Our operators ${\cal O}_d$ and $e_d$ likewise do not depend on $N$, and should 
thus be thought of as representations of the respective generators of the universal 
symmetric functions ring.
This is one of the benefits of the second quantized approach developed here. Note that indeed the
$e_d$ automatically annihilate any state of $N<d$ particles.

\subsection{Partitions and dominance}

We may expand general many-body states in the lowest Landau level 
as
\be\label{expansion}
  \ket{\psi}= \sum_\lambda C_\lambda \ket{\lambda}\,,
\ee
where the $\ket{\lambda}$ denote a basis of occupation number eigenstates,
with respect to an appropriate lowest Landau level basis of single particle states.
The latter will always be chosen as eigenstates of some guiding-center or momentum
quantum number. One way to denote a ket $\ket{\lambda}$ is therefore through a
partition $\ell_1\geq\ell_2\geq\dotsc\geq\ell_N$ for bosons, or 
$\ell_1 >\ell_2 >\dotsc >\ell_N$ for fermions, where $\ell_i$ denotes the orbital index
of the $i$th particle. Since the Hamiltonian conserves momentum, when discussing zero modes
we may assume that
\be
   L=\ell_1+\dotsc +\ell_N
\ee
is the same for all kets $\ket{\lambda}$ contribution to \Eq{expansion}.
We will also refer to the kets $\ket{\lambda}$ as partitions of the number $L$.
There is  a standard notion of ``dominance'' for partitions whose utility
in the context of quantum Hall states has been noted previously.\cite{bernevig1, bernevig2}
We say that $\klambda$ dominates $\klambdap$, or $\klambda \geq \klambdap$, 
if $\klambda=\klambdap$ or $\klambdap$ can be generated from $\klambda$ (up to 
normalization factors or phases) by repeated application
of inward ``squeezing operations'':\cite{RH}
\be
       c^\dagger_{r_1} c^\dagger_{r_2} c_{r_2+d}^{\;} c_{r_1-d}^{\;},\quad r_1 \leq r_2,\; d>0\,.
\ee
An equivalent, and technically more useful characterization of dominance is
given by
\be\label{domdef}
\begin{split}
 &\{\ell_i\}_{i=1\dotsc N} \geq  \{\ell'_i\}_{i=1\dotsc N} \\
 \Leftrightarrow &\;\; \sum_{i=1}^n \ell_i \geq \sum_{i=1}^n \ell'_i \quad\mbox{for all $n=1\dotsc N$.} 
\end{split}
\ee
We we always refer to this definition here.

A  special role is further played by partitions satisfying a ``generalized Pauli 
principle'':\cite{bernevig1, bernevig2} Here, we will say that a partition $\klambda$ 
satisfies the $M$-Pauli-principle if no more
than $1$ particle is present in any $M$-consecutive orbitals, or $\ell_i\geq \ell_{i-1}+M$.
In particular, the state \eqref{dominance} satisfies the $M$-Pauli principle, and is the 
lowest-$L$ state that does so for given particle number $N$.
We will also say that a general superposition \eqref{expansion} is dominated by a partition 
$\ket{\lambda_0}$ if $C_{\lambda_0}\neq 0$ and $\ket{\lambda_0}\geq \klambda$ for every
$\lambda$ with $C_\lambda \neq 0$. 
Note that we require 
that the dominant partition  $\ket{\lambda_0}$ appears with nonzero coefficient also, which is
not always the case in the literature.
Thus, in particular, the statement made in \Eq{dominance} and below is that
$\ket{\psi_{1/M}}$ is dominated by $\ket{\tilde\psi_{1/M}}$. 

Our strategy for proving Theorem 1 is now the following. It will turn out that, using general results 
of Ref. \onlinecite{ortiz}, Theorem 1 can be obtained as a corollary of the following:

{\em
{\bf Theorem 2}. 

For any partition $\klambda$ satisfying the $M$-Pauli-principle, there is a zero mode $\ket{\psi_\lambda}$ that 
is dominated by $\klambda$ and that is of the form
\be\label{T2}
   \ket{\psi_\lambda} = \prod_{\alpha=1}^k e_{d_\alpha} \ket{\psi_{1/M}}\,.
\ee
}
In particular, $\ket{\psi_\lambda}$ is obtained by applying a special element of the algebra $\cal A$
to the ``incompressible'' zero mode $\ket{\psi_{1/M}}$.
 Note again that the order of operators in \Eq{T2} does not matter.
 
\section{Proofs\label{proofs}}

We will prove Theorem 2 first.
Observe that the action of the operator $e_d$ on the partition $\klambda$
is just to promote the orbital indices of $d$ particles by $1$, in all possible ways.
It is thus useful to consider the following decomposition of the operator $e_d$,
where we consider fermions first:
\be
e_d=\sum_{\substack{S\subset 2^{\bar N}\\ |S|=d}} \mathfrak{e}_S\quad\mbox{(fermions).}
\ee 
Here, for an integer $N$, the bar denotes the set $\bar N=\{1,\dotsc,N\}$, 
 $2^{\bar N}$ denotes the set of all subsets of $\bar N$, and the sum goes over all
 such subsets of $d$ elements. $\mathfrak{e}_S$ then is an operator that acts of the partition $\klambda$
 by promoting the particles corresponding to the subset $S$, where this correspondence 
 is established by ordering particles according to their orbital index in $\klambda$.
 Hence, if $\klambda$ corresponds to the partition $\{\ell_i\}_{i=1\dotsc N}$ and $\mathfrak{e}_S\klambda$
 is a partition made from the numbers $\ell_i$ for $i\notin S$ and $\ell_{i}+1$ for $i\in S$,
 and for fermions, the state will be annihilated if this would lead to double occupancies.
 We may choose phase conventions for the $\klambda$-basis such that $\ek_S\klambda$
 is always either zero or equal to another basis state $\klambdap$.
 
 In particular,  we may write
 \begin{subequations}\label{fermbos}
 \be\label{ferms}
e_d=\mathfrak{e}_{\bar d}+ \sum_{\substack{S\subset 2^{\bar N}\\ |S|=d\\S\neq \bar d}} 
\mathfrak{e}_S\quad\mbox{(fermions),}
\ee 
where $\mathfrak{e}_{\bar d}$ is an operator that promotes the $d$ particles of highest 
orbital index, by one orbital index (and never annihilates a partition $\klambda$). Note that the
operators $\mathfrak{e}_{\bar d}$ commute, just as the $e_d$ do.

The situation is not really more complicated for bosons, but for accuracy, we should write
\be\label{bosons}
e_d=\mathfrak{e}_{\bar d} b_{\bar d}+ \sum_{\substack{S\subset 2^{\bar N}\\ |S|=d\\S\neq \bar d}} 
\mathfrak{e}_S b_S\,, \quad\mbox{(bosons),}
\ee
\end{subequations}
where the $b_S$ are positive operators that act diagonally on the basis of occupation number
eigenstates, and multiply state with multiple occupancies by necessary combinatorial factors.
These factors are necessary, since we still insist that $\ek_S$ acting on $\klambda$ gives
another basis state $\klambdap$, with unit coefficient (no annihilation occurs for bosons).
When acting on a $\klambda$ with multiple occupancies, there are different subsets $S$ that have
the same effect on $\klambda$, and are then associated with the same terms in \Eq{elementary2}. 
The resulting ambiguity in the operators $b_S$ can be resolved arbitrarily. The point is not worth further 
elaborating, since no such ambiguities exist whenever $\ek_{\bar d}$ acts on a state $\klambda$ without double
occupancies, in which case we always have $b_{\bar d}\klambda =\klambda$. This is in particular 
the case for any $\klambda$ satisfying the $M$-Pauli-principle. We may thus proceed without 
distinguishing between fermions and bosons.

Theorem 2 now follows from the following simple facts:
\begin{subequations}\label{simplefacts}
\be\label{fact1}
\klambda\geq\klambdap \Leftrightarrow \ek_{\bar d} \klambda \geq \ek_{\bar d} \klambdap
\ee
\be\label{fact2}
   \ek_{\bar d} \klambda \geq \ek_S \klambda \quad\mbox{for $|S|=d$.}
\ee
Note that by definition of the operators $\ek_S$, both sides of \Eq{fact2} are basis
states and can be identified with partitions, such that the relation is meaningful, {\em except}
where the right hand side vanishes, in which case we take the relation to be satisfied by convention.
The simple proofs are relegated to Appendix \ref{dominanceApp}.
\end{subequations}
Useful consequences of Eqs. \eqref{simplefacts} are
\be\label{fact3}
\klambda\geq\klambdap \Rightarrow \ek_{\bar d} \klambda \geq \ek_{S} \klambdap
\quad \mbox{for $|S|=d$,}
\ee
which uses the transitivity of the dominance relation,
and
\be\begin{split}\label{fact4}
   \mbox{If}\;\;\klambda\geq\klambdap\;\;\mbox{and} \;\;
   0\neq \ek_{S} \klambdap \geq \ek_{\bar d} \klambda\,, \;\;|S|=d\,,\\ \mbox{then}\quad \klambda=
   \klambdap\;\;\mbox{and}\;\; \ek_{S} \klambdap = \ek_{\bar d} \klambda \,.
\end{split}
\ee
The last follows from \Eq{fact3}, the transitivity and anti-symmetry of the
dominance relation, and the `$\Leftarrow$' direction of \Eq{fact1}. 

From these facts, and the knowledge that $\psiM$ is dominated by $\tpsiM$,
one immediately obtains that the right hand side of \Eq{T2}
is dominated by the first term on the right hand side of the following
\be\label{compartments}
\prod_{\alpha=1}^k e_{d_\alpha} \ket{\psi_{1/M}}=\prod_{\alpha=1}^k \ek_{\widebar{d_\alpha}} 
\ket{\tilde\psi_{1/M}} + \mbox{subdominant.}
\ee
The proof is by simple induction in $k$, and uses only Eqs. \eqref{fact3}, \eqref{fact4}.
Suffice it to give details for $k=1$. We act with \Eq{ferms} or \Eq{bosons}, whichever applies, on
\Eq{dominance}. Since the first term in \Eq{dominance} dominates all others,
\Eq{fact3} immediately implies that every term generated in the action of 
\Eq{fermbos} on \Eq{dominance} is dominated by the first term in \Eq{compartments}.
We must also make sure that contributions to this term cannot cancel.
By \Eq{fact4}, they cannot be canceled by any contributions coming from the
sum in \Eq{dominance}. However, contributions from the first term cannot cancel either,
since, 
for any $\klambda$ without double occupancies, it is easy to see
that different operators $\ek_S$ generate different states $\ek_S\klambda$. In particular,
that is the case if $\klambda$ satisfies the $M$-Pauli principle, as the first term in
\Eq{dominance} or in \Eq{compartments} (see below) does.

Now, suppose $\klambda$ is a partition that satisfies the $M$-Pauli-principle. 
Then $\ek_{\bar d}$ acts on $\klambda$ by squeezing in another zero to the 
left of the particle that is the $d$th particle from the right, by means of right-pushing 
this particle and every particle to its right by one orbital. We can think of any 
$\klambda$ satisfying the $M$-Pauli-principle as consisting of $N$ compartments 
of zeros separated by $1$s, where the $d$th compartment is to the left of the $d$th 
particle, counted right to left, and has at least $M-1$ zeros, except for $d=N$th 
compartment, which may be devoid of zeros.
Then the operator $\ek_{\bar d}$ just fills another zero into the $d$th compartment.
Given that $\tpsiM$ is the densest partition satisfying the
$M$-Pauli principle (all compartments have the minimum possible number of zeros),
it is clear that every $\klambda$ satisfying the $M$-Pauli-principle can be written in the
form displayed by the first term on the right hand side of \Eq{compartments}.
This completes the proof of Theorem 2.

For Theorem 1, we now consider the operator 
\be
    S=\sum_{r\geq 0} r^2 c^\dagger_r c_r\,.
\ee 
When working in the cylinder geometry, this can be thought of as the generator of changes
in the cylinder's radius.\cite{zhou12}
It is easy to see that all partition states $\klambda$ are eigenstates of $S$, and that inward squeezing
always lowers the value of $S$. So, if by $S_\lambda$ we denote the $S$-eigenvalue of 
$\klambda$, $\klambda\geq\klambdap$ implies $S_\lambda\geq S_{\lambda'}$, with 
equality in the latter only for equality in the former.
For an arbitrary ket $\ket{\psi}$, we denote by $S(\psi)$ the largest $S$-value of any partition
contributing to the expansion \eqref{expansion}  of the state. Hence if $\ket{\psi}$ is dominated by
$\lambda$, then $S(\psi)=S_\lambda$. It is for this reason that the dominant partition also
coincides with the thin cylinder limit of the state.

We now fix the particle number $N$, and prove Theorem 1 by induction over the possible 
values of $S(\psi)$ for zero modes $\ket{\psi}$.
The general results of Ref. \onlinecite{ortiz} show that every partition $\klambdap$ contributing to
a zero mode $\ket{\psi}$ can be obtained from a partition $\klambda$ (not necessarily always 
the same) via inward squeezing, where $\klambda$ contributes to $\psi$ {\em and} satisfies 
the $M$-Pauli-principle. 
Then, by Theorem 2, the possible values $S(\psi)$ for zero modes are exactly
the values $S_\lambda$ for $\klambda$ satisfying the $M$-Pauli-principle.
The lowest such $S_\lambda$ is uniquely obtained for $\klambda=\tpsiM$.
A corresponding zero mode with $S(\psi)=S_{\tilde\psi_{1/M}}$ must then be dominated by 
$\tpsiM$, and the unique zero mode for which this is true is denoted $\psiM$ in \Eq{dominance}.
The uniqueness can be followed from the fact that if two different zero modes were dominated
by $\tpsiM$, we could make a non-trivial linear combination that is also a zero mode 
and is not dominated by any partition
satisfying the $M$-Pauli-principle, in violation of the general rules for zero modes found in Ref. \onlinecite{ortiz}.
Thus, for the smallest possible $S$-value $S(\psi)=S_{\tilde\psi_{1/M}}$, the statement of Theorem 1 is clearly correct.

The induction step proceeds similarly, but makes further use of Theorem 2.
Suppose Theorem 1 has been shown for all zero modes $\ket{\psi}$ with $S(\psi)<S$, for 
some $S$. Now consider a zero mode $\ket{\psi}$ with $S(\psi)=S$.
Then we can write
\be\label{psiexpansion}
\ket{\psi}= \sum_{i=1}^n a_i \ket{\lambda_i} +\sum_{\lambda'}C_{\lambda'} \klambdap\,,
\ee
with $S_{\lambda_i}=S$, $a_i\neq 0$, and for every $\klambdap$ with $C_{\lambda'}\neq 0$,
$S_{\lambda'}<S$. Invoking again the aforementioned general results\cite{ortiz}
the $\ket{\lambda_i}$ must each be dominated by a partition that satisfies the $M$-Pauli-principle,
and that has non-zero coefficient in \Eq{psiexpansion}. However, no term in the second 
sum can dominate any term in the first, since the latter have larger $S$-value. It follows, then, that all the 
$\lambda_i$ must themselves satisfy the $M$-Pauli principle. If we now choose zero modes
$\ket{\psi_{\lambda_i}}$ as in Theorem 2, then
\be\label{finalmode}
    \ket{\psi}-\sum_{i=1}^n a_i \ket{\psi_{\lambda_i}}
\ee
is a zero mode, and the coefficients of the partitions $\ket{\lambda_i}$ cancel.
Then all partitions $\klambdap$ contributing to \Eq{finalmode} have $S_{\lambda'}$ less than our
given $S$. This is so since it is true for every term in the second sum of \Eq{psiexpansion},
and also for every partition contributing to $\ket{\psi_{\lambda_i}}$ {\em except} $\ket{\lambda_i}$,
since  $\ket{\psi_{\lambda_i}}$ is dominated by $\ket{\lambda_i}$.
By induction, the statement of Theorem 1 then applies to the zero mode \eqref{finalmode}.
But then it also applies to $\ket{\psi}$, since it does apply to each $\ket{\psi_{\lambda_i}}$ 
individually, according to \Eq{T2}.

The same inductive procedure in $S$ can be used to show that the states $\ket{\psi_\lambda}$
defined in Theorem 2 are linearly independent, and are therefore a basis of all zero modes.
This establishes the well-known one-to-one correspondence between Laughlin-like zero modes
and patterns satisfying the $M$-Pauli principle, which was first obtained by thin 
cylinder\cite{RH,seidel05,BK2} methods as well as Jack polynomial methods 
and its generalization to fermions\cite{bernevig1, bernevig2, regnault} (see also Ref. \onlinecite{Read06}).
Note that while $\klambda$ dominates $\ket{\psi_\lambda}$ defined in \Eq{T2}, it is not necessarily
the only partition satisfying the $M$-Pauli-principle contributing to this state. By forming
new linear combinations, we may define a new zero mode basis $\ket{\phi_\lambda}$ such that
$\ket{\lambda}$ is the only partition satisfying the $M$-Pauli-principle contributing to $\klambda$.
This requirement uniquely defines $\ket{\phi_\lambda}$, up to normalization, by an 
argument analogous to that of the uniqueness of a zero mode dominated by $\tpsiM$ 
given above. To show the existence of such a basis $\ket{\phi_\lambda}$, one again 
uses Theorem 2 and induction in $S$.
We thus note the following corollary to Theorem 2:

{\em
{\bf Theorem 2'}. 

For any partition $\klambda$ satisfying the $M$-Pauli-principle, there is a state
\be
\ket{\phi_\lambda}= {\hat a}\psiM\,,\quad \mbox{with ${\hat a}\in{\cal A}$},
\ee
which is dominated by $\klambda$, with the additional property that $\langle \lambda'|\phi_\lambda\rangle =0$
for every $\klambdap$ satisfying the $M$-Pauli principle other than $\klambda$.
}
\\

Finally, the linear independence of the states in \Eq{T2} shows that the $e_d$ 
generate the commutative algebra $\cal A$ freely. This is not surprising, given the 
relation with polynomial rings that we know of, but did not need to make use of so far. 

\section{Relation with bosonization\label{boson}}
We now recall that our original motivation was to answer questions about the operators
${\cal O}_d$, \Eq{Od}, in particular the question of the completeness of the zero modes they generate
when acting on $\psiM$. We have answered this question positively.
The ${\cal O}_d$ are in some sense more interesting physically than their counterparts $e_d$,
since they are single particle operators. One cannot help but noticing that, especially 
with the ``thick cylinder'' normalization convention chosen here,\footnote{additional 
normalization factors appearing for other geometries are obtained as explained in 
Sec. \ref{OP}.} operators similar to \eqref{Od} appear in every bosonization 
dictionary (even though the operators $c_r^{\;}$, $c_r^\dagger$ create fermions only for 
$M$ odd). They represent density modes, which in the quantum Hall context are naturally 
associated with edge excitations.\cite{wen_edge} The one-dimensional edge theory 
of a two-dimensional bulk system is necessarily a {\it long-wavelength effective theory}, 
since at higher quantum numbers, the excitations penetrate deeper and deeper into 
the bulk, removing themselves from the edge (and are, for quantum Hall states, more 
properly thought of as quasi-holes). It is thus remarkable that in the present context, 
the operators \eqref{Od} appear as generators of exact eigenstates of a microscopic 
bulk theory (Loop-algebraic Hamiltonian) that does not require any thermodynamic 
limit to be taken. Similar observations 
can of course be made at the level of polynomial wave functions,\cite{haldane_prange,RR96, stone_schur} 
though we regard it as additional benefit
that ``bosonic mode operators'' of the form \eqref{Od} can be given an exact meaning, 
with their relation to the microscopic electron operator absolutely explicit, without the need 
to take any large $N$ limit or apply a {\it normal ordering} operation. This, in our opinion, makes the 
correspondence between edge 
and bulk physics of the Laughlin states particularly lucid. 

As one application of our results to the bosonization method, recall that every 
bosonization scheme needs to address the equivalence of the fermionic and 
bosonic Hilbert spaces. (Notice, however, that 
the bosonization scheme we are referring to is performed on an angular momentum lattice [not a 
real space lattice],  and for $M>1$ the Hilbert space equivalence applies only to the  zero mode subspace $Z$ 
of the total Hilbert space.)  
Usually, this is done either through partition functions,\cite{haldane_bos} or 
through Schur functions\cite{stone_schur}.
If we apply the results of the preceding sections to the special case $M=1$, we obtain an alternative
proof of this statement. In this case, the $M$-Pauli-principle is just the standard 
Pauli principle for fermions. Every fermionic occupancy eigenstate $\klambda$ 
satisfies the $M=1$ Pauli principle.
Consistent with that, the Hamiltonian \eqref{pseudo} vanishes identically in this case, and the 
subspace of zero modes $Z$ is really the entire fermionic Fock space.
Indeed, in the light of the above the statement of Theorem 2' just becomes, in this 
special case, that every $N$-particle fermionic occupation number eigenstate 
$\klambda$ can be written as the $N$-particle ``vacuum'' state 
$\ket{\psi_{M=1}}=\ket{1111\dotsc 100000\dotsc}$, acted upon by a linear combination 
of products of operators ${\cal O}_d$. For, since we can drop the qualifying statement
``satisfying the $M$-Pauli-principle'' for $M=1$, it follows that $\ket{\phi_\lambda}=\klambda$.
This is just the statement that the subspace generated by the operators ${\cal O}_d$ 
when acting on the $N$-particle ``vacuum'' is just the entire $N$-particle sector 
of the fermionic Fock space.
Note that some authors like to set up the bosonization scheme starting 
with large but finite $N$,
taking the limit $N\rightarrow \infty$ only at the end.\cite{schonhammer}
It is then useful to know that the identity between the two Hilbert spaces is already 
exact for finite $N$. This is not obvious from the usual partition function 
technique.\cite{haldane_bos} It does, however, also follow from Stone's Schur function 
method.\cite{stone_schur}
Our approach is more naturally related to the latter, although we avoid the 
language and apparatus of symmetric polynomial rings (with the exception 
of the Newton-Girard operator relations, which, nevertheless, do not make explicit 
reference to polynomials). In its general, $M\geq 1$ form, however, our 
approach gives an exact treatment of zero modes in the correlated Laughlin 
states  while making the relation between generators of such zero modes 
and the second quantized electron operator explicit.
Another direct application of this explicit relation is given in the following section.
\section{Relation to the string order parameter\label{OP}}

A problem of great conceptual importance is to understand the relationship 
between the topological order in fractional quantum Hall liquids and more conventional
orders. In addition, it is desirable to characterize this type of hidden order to 
embed it in the more general context of topological orders in quantum matter, 
probably helping to unveil a so elusive classification. 
Much insight has been drawn in particular from parallels with the off-diagonal
long range order in superfluids, where, in the quantum Hall context, the
corresponding order parameter is necessarily non-local. In the approach 
pioneered by Read,\cite{readOP} such non-locality comes about by multiplying
the local electron creation operator with powers of the non-local ``quasi-hole operator''.
In the preceding sections, we have avoided any contact between our second quantized formalism
and the language of first-quantized analytic wave functions.
In this section, we breach this barrier, in order to give another application of our results.
The quasi-hole operator can be defined, for the disk geometry, as the operator
multiplying wave functions in the lowest Landau level by the function\cite{Laughlin}
\be\label{U}
U(\zeta) = \prod_{i=1}^N (z_i-\zeta)\,.
\ee
Here, $z_i=x_i+ {\rm i} y_i$ ($\bar{z}_i=x_i- {\rm i} y_i$) represents the position 
of the electron $i$ in the complex plane, and 
$\zeta$ is the  location of the quasi-hole in complex-variable notation.

The problem of giving a second quantized form to this operator has been 
posed and analyzed by Stone.\cite{stone_book}
There, the operator corresponding to \Eq{U} is defined abstractly, through its intertwining relation 
with the electron field creation operator $\psi(z)^\dagger$,\footnote{Here and in the following, 
$\psi(z)$ is always short for $\psi(z,\bar z)$, and similarly for $\phi_r(z)$.}
\begin{subequations}\label{axiomatic}
\be\label{Upsi}
  \hat{U}(\zeta) \psi(z)^\dagger =(2\partial_{\bar z} +\frac 12 z-\zeta)\psi(z)^\dagger \hat{U}(\zeta) \quad\mbox{(disk)}
\ee
valid for the disk geometry,
and the additional requirement
\be\label{ax2}
  \hat{U}(\zeta)\ket{0}=\ket{0}\,.
\ee
\end{subequations}
In the equation above $2 \partial_{\bar z}=\partial_x + {\rm i} \partial_y$. 
Equations \eqref{axiomatic} allow one to work out the effect of the operator $\hat{U}(\zeta)$
on arbitrary states created out of the vacuum by means of products of electron creation
operators. It can be argued\cite{stone_book} that this action agrees with that defined 
by the multiplication of first quantized wave functions with the function given in \Eq{U}. 
However, the relation of the operator
$\hat{U}(\zeta)$ with the second quantized electron creation operator remains somewhat obscure.
Here we specify this relation explicitly, which is needed in order to specify 
Read's non-local (string) order parameter $\hat{K}(\zeta)$.

It is easy to multiply out the product in \Eq{U}, giving

\be\label{U2}
  U(\zeta)= \sum_{d=0}^N (-\zeta)^{N-d} s_d \,,
\ee
where $s_d$ is the elementary symmetric polynomial defined in \Eq{elementary1}.
In the above, we have identified explicitly the second quantized operators that facilitate
multiplication with elementary symmetric polynomials, and expressed them through electron
creation and annihilation operators. This gives
\be\label{U3}
     \hat{U}(\zeta)= \sum_{d=0}^N (-\zeta)^{N-d} e_d \,,
\ee
 where the $e_d$ are inferred from \Eq{elementary2}, except that we must now work
with the normalization conventions of the disk geometry.
This corresponds to applying a similarity transformation $e_d\rightarrow e^\mS e_d e^{-\mS}$
to all of the operators, which is defined via\cite{ortiz} 
\be
  e^\mS c_r e^{-\mS} = {\cal N}_r^{-1} c_r,\quad  e^\mS c_r^\dagger e^{-\mS} = {\cal N}_r c_r^\dagger\,,
\ee
and ${\cal N}_r$ is a factor related to the normalization constant of single particle orbitals
for the geometry in question.
For the disk geometry, we have
${\cal N}_r = \sqrt{2\pi 2^r r!}$.
In this section, by $e_d$ we will always mean the operators that have undergone
the appropriate similarity transformation in question (and continue to represent the
multiplication with elementary symmetric polynomials in the first quantized picture 
of this geometry). We denote the original operators defined in \Eq{elementary2} by
$E_d$, which correspond to the geometry of an infinitely thick cylinder.
The $E_d$ satisfy the commutation relation
\be\label{ecomm}
  [E_d, c^\dagger_r]= c^\dagger_{r+1} E_{d-1}\,,
\ee
which, after conjugation with $e^\mS$, becomes
\be\label{ecommgen}
 [e_d, c^\dagger_r]= \frac{\cN_{r+1}}{\cN_r}c^\dagger_{r+1} e_{d-1}\,,
\ee
or, for the disk geometry
\be\label{commdisk}
 [e_d, c^\dagger_r]= \sqrt{2r+2}\, c^\dagger_{r+1} e_{d-1}\quad\mbox{(disk).}
\ee

It is advantageous to rewrite \Eq{U3} as
\be \label{Ufinal}
   \hat{U}(\zeta)= (-\zeta)^{\hat N} \sum_{d=0}^\infty (-\zeta)^{-d} e_d \,,
\ee
where $\hat N$ is the particle number operator, and we have used the fact that 
$e_d$ annihilates states with particle number $N<d$.
Hence the operator depends explicitly on particle number only through the trivial pre-factor.
The latter is not all that important when working within a subspace
of constant particle number, but is crucial in the following intertwining relation,
\be\label{Ucdisk}
  \hat{U}(\zeta) c^\dagger_r = -\zeta c^\dagger_r \, \hat{U}(\zeta) + \sqrt{2r+2} \,
  c^\dagger_{r+1} \, \hat{U}(\zeta)\quad\mbox{(disk),} 
\ee
which follows straightforwardly from Eqs. \eqref{commdisk} and \eqref{Ufinal}.
\Eq{Ucdisk} allows us to show directly 
that the second quantized quasi-hole operator, which we have
explicitly defined in terms of the electron operator in \Eq{Ufinal}, does indeed
have the property \eqref{Upsi}, as originally conjectured by Stone.\cite{stone_book} 
Here we prove relation \eqref{Upsi} and extend it to arbitrary geometries. 
To this end, we write out the mode expansion of the field operator 
$\psi^\dagger(z)$ in terms of $c^\dagger_r$s:\cite{Note2}
\be
     \psi^\dagger(z)= \sum_{r=0}^\infty \phi_r^\ast(z) c^\dagger_r\,,
\ee 
where, for the disk, the single particle orbitals are
\be
   \phi_r (z)= \frac{1}{\sqrt{2\pi 2^r r!}} z^r e^{-|z|^2/4} \quad\mbox{(disk).}
\ee
With the use of \Eq{Ucdisk}, one obtains
\be\label{Ucgeneral}
\hat{U}(\zeta) \psi^\dagger(z) = -\zeta \psi^\dagger(z) \hat{U}(\zeta) + 
\sum_{r=1}^\infty \frac{\cN_{r}}{\cN_{r-1}} \phi^\ast_{r-1} (z) c^\dagger_r \
{ \hat{U}(\zeta)}\,,   
\ee
where we have again identified  $\sqrt{2r}$ with $\frac{\cN_{r}}{\cN_{r-1}}$ for greater generality.
We now note that indeed \Eq{Upsi} follows simply from the fact that
\be
      D=2\partial_{\bar z} +\frac 12 z\quad\mbox{(disk)}
\ee
is a differential operator that satisfies
\begin{subequations}
\label{properties}
\be\label{kill0}
D\phi^\ast_0(z) =0
\ee
and
\be\label{shift}
  D\phi^\ast_r(z)=\frac{\cN_{r}}{\cN_{r-1}} \phi_{r-1}^\ast(z)\quad\mbox{for $r>0$.}
\ee

\end{subequations}

We also note that the operator \eqref{Ufinal} trivially satisfies \Eq{ax2},
which completes the demonstration that our expression \eqref{Ufinal}
satisfies the axiomatic properties of Ref. \onlinecite{stone_book}.

As a further application, we generalize Eqs. \eqref{Upsi}
and \eqref{Ucdisk} to the cylinder and sphere geometries.
We begin with \Eq{Ucdisk} which is useful in practical computations.
This is trivial, since all we have to do is to identify the factor 
$\sqrt{2r+2}$ once again with the ratio $\cN_{r+1}/\cN_r$ of normalization
constants of the corresponding single particle orbitals.
For the cylinder of finite radius $R_y=1/\kappa$, we use the standard (Landau) gauge
where
\be
  \phi_r(z)= (4\pi^3)^{-1/4}\sqrt{\kappa}e^{-\frac 12 \kappa^2 r^2} \xi^r e^{-\frac 12 x^2} \;\;\mbox{(cylinder),}
\ee
with $\xi=e^{\kappa z}$ playing the role of the variable in the polynomial part of wave functions.
The constant $\cN_r$ is the inverse of the factor multiplying the monomial (here $\xi^r$),\cite{ortiz}
where $r$-independent factors can be dropped as only ratios of these constants matter in the following.
We may hence choose $\cN_r=\exp(\frac 12 \kappa^2 r^2)$ for the cylinder.
This then yields
\be\label{Ucyl}
  \hat{U}(\zeta) c^\dagger_r = -\zeta c^\dagger_r\, \hat{U}(\zeta) + e^{\kappa^2(r+\frac 12)}c^\dagger_{r+1} 
  \hat{U}(\zeta)\quad\mbox{(cylinder),} 
\ee
where, for the cylinder, $\zeta$ is related to the complex quasi-hole coordinate as $\xi$ is to $z$.

For the spherical geometry, we follow Ref. \onlinecite{RR96} in identifying  a sphere of 
radius $R$ with the infinite plane with
complex coordinate $z$ via stereographic projection.
As described there, this leads to normalized single particle orbitals of the form
\be\label{phisphere}
\begin{split}
  \phi_r(z)=\frac{1}{(2R)^{1+r} }\sqrt{\frac{N_\Phi+1}{4\pi}}{N_\Phi \choose r}^{\frac 12} 
  \frac{z^r}{(1+\frac{|z|^2}{4R^2})^{1+\frac{N_\Phi}{2}}}\\
  r=0\dotsc N_\Phi\mbox{    (sphere),}
  \end{split}
\ee
with $N_\Phi$ being the number of magnetic flux quanta piercing the sphere.
Following the same reasoning as for the cylinder,  
 this leads to $\cN_r=(2R)^{1+r}{N_\Phi\choose r}^{-\frac 12} $, giving
 \be\label{Usph}
  \hat{U}(\zeta) c^\dagger_r = -\zeta c^\dagger_r\, \hat{U}(\zeta) + 2R\sqrt{\frac{r+1}{N_\Phi-r}}c^\dagger_{r+1} 
  \hat{U}(\zeta)\quad\mbox{(sphere).} 
\ee
 
 It may be of some interest to also generalize \Eq{Upsi} to the other geometries, 
 the starting point being the general \Eq{Ucgeneral}.
 We will see that the local character of \Eq{Upsi} is somewhat lost 
 for a finite-radius cylinder or sphere, and is recovered
 only in the large-radius limit.
 The generality of the discussion following \Eq{Ucgeneral} implies that
 all we need to do is to find a differential operator $D$ that satisfies Eqs. \eqref{properties},
 for the appropriate orbitals and normalization factors.
 We then have
 
 \be\label{Upsigen}
  \hat{U}(\zeta) \psi(z)^\dagger =(D-\zeta)\psi(z)^\dagger \hat{U}(\zeta) \;\mbox{.}
\ee
For the cylinder, it is more convenient to let the orbital index $r$ roam
over all integers, including negative. This leaves the relevant commutation relations
intact. (One only loses the equivalence between the zero mode counting of the disk
and cylinder geometries, which mattered in Sec. \ref{proofs} but not in the present
context.) Thus, for the cylinder, we abandon \Eq{kill0} and only seek to enforce \Eq{shift}. 
The latter essentially says that $D$ must be
a magnetic translation (as the orbitals magnetically translate  into one another on the cylinder),
except for the trivial normalization factor. 
Alternatively, it is elementary to see that $D=e^{-\kappa(\bar z +\kappa)} 
e^{\kappa(2\partial_{\bar z} +\frac z 2 + \frac{\bar z}{2})}$ has all desired properties, 
where the derivative produces a factor of $e^{2r\kappa^2}$ when acting on $\phi_r$, 
and the following term $\frac z 2 + \frac{\bar z}{2}$ only serves to cancel the derivative-action
on $e^{-\frac 12 x^2}$. Thus we have
 \be\label{Upsicyl}
 \begin{split}
  \hat{U}(\zeta) \psi(z)^\dagger =(e^{-\kappa(\bar z +\kappa)} e^{\kappa(2\partial_{\bar z} +
  \frac z 2 + \frac{\bar z}{2})}-\zeta)\psi(z)^\dagger \hat{U}(\zeta) \\
  \mbox{(cylinder).}
\end{split}
\ee
 
For the sphere, rather than dropping the orbital cutoff at $r=0$ as for the cylinder, we must enforce
an additional cutoff at $r=N_\Phi$, due to the finite dimensionality of the Hilbert space.
We may do so by using the convention
\be
   c_r^{\;}=c_r^\dagger=0 \quad\mbox{for $r>N_\Phi$.}
\ee
One may see that the commutation relation \eqref{ecommgen}
then still holds even for $r=N_\Phi$.
Then we may seek an operator $D$ satisfying Eqs. \eqref{properties},
and note that the operator
\be
 \Delta=\frac{4R^2}{N_\Phi} \left(\partial_{\bar z} +\frac{z}{4R^2} (1+\frac{|z|^2}{4R^2})^{-1}(1+\frac {N_\Phi} 2)\right) 
\ee 
 satisfies \Eq{kill0} and also \Eq{shift} for the special case of $r=1$.
 Again, the terms following the derivative only serve to cancel its action on
 the non-holomorphic part in \Eq{phisphere}.
 Thus, $\Delta$ also satisfies
 \be\begin{split}
    \Delta^ m \phi_r^\ast(z) &= (\frac{4R^{2}}{N_\Phi})^{m}m!{r \choose m} \bar z ^{-m} \phi_r^\ast(z)\\
   &= \frac{(2R)^{2m-1}}{N_\Phi^m} m!{r \choose m} \sqrt{\frac{N_\Phi-r+1}{r}}\bar z ^{-(m-1)} \phi_{r-1}^\ast(z)\\
    &\qquad\qquad\qquad\qquad\qquad\qquad\qquad\qquad \mbox{(sphere).}
    \end{split}
 \ee
 In particular, $ \Delta^ m \phi_r^\ast(z)=0$ for $m>r$.
 These observations motivate the following Ansatz:
 \be\label{Dsphere}
     D=\sum_{r=1}^{N_\Phi} a_r\, \bar z ^{r-1} \Delta ^r\quad \mbox{(sphere)}
 \ee
with $a_1=1$.
 Plugging this into \Eq{shift} for each $r$ leads to the relations
 \be\label{recursive}
 (N_\Phi-r+1)(r-1)! \sum_{m=1}^r \frac{1}{(r-m)!} \frac{(2R)^{2m-2}}{N_\Phi^m} a_m = 1\,,
 \ee
 from which the $a_r$ may be determined recursively. It turns out that the unique solution
 to these equations can be given as
 \be\label{ar}
    a_r= \frac{N_\Phi^r}{(2R)^{2(r-1)}}\frac{(N_\Phi-r)!}{N_\Phi!}\,.
 \ee
Equation \eqref{Upsigen},  along with Eqs. \eqref{Dsphere} and \eqref{ar}, then generalize
 \Eq{Upsi} to the sphere.
 It is clear from \Eq{ar} that in the limit of a large sphere, higher derivatives
 are unimportant in \Eq{Dsphere}. In particular, in the limit $N_\Phi\rightarrow \infty$
 with $N_\Phi/R^2$ fixed and equal to $2$, one recovers the equation derived for the disk,
 \Eq{Upsi}, as expected.\cite{RR96}
 
 In closing this section, we make two remarks. Notice, first, that the state 
 \be
 \hat{U}(\zeta) \ket{\psi_{1/M}}
 \ee
 is also a zero mode of a Haldane pseudo-potential-type Hamiltonian. The 
 quasi-hole operator $\hat{U}(\zeta)$ preserves the number of particles $N$ but changes the 
 number of fluxes of the incompressible state $\ket{\psi_{1/M}}$. Second, our 
approach allows us to explicitly express Read's non-local 
 (string) order parameter
 \be
\hat{K}(\zeta)= \psi(\zeta)^\dagger \hat{U}(\zeta)^M
 \ee
in  second-quantized form, and thus can be used to study off-diagonal long-range 
order in Haldane pseudo-potential-type systems
\be
\lim_{|\zeta-\zeta'| \rightarrow \infty} 
\langle \psi_{1/M}| \hat{K}(\zeta)^\dagger \hat{K}(\zeta')\ket{\psi_{1/M}} \rightarrow {\sf constant} \ .
 \ee

  \section{Discussion and Conclusion}
 
 In this paper, we have developed a representation of the ring of symmetric functions
 using bosonic or fermionic ladder operators satisfying canonical commutation
 relations. In particular, one set of operators we constructed can be understood as
affecting states by multiplying associated first quantized wave functions in the lowest Landau level
with elementary symmetric polynomials.
Our primary motivation for constructing this representation is to provide tools for an 
alternative, polynomial-free or second quantized approach to represent physics in 
the lowest Landau level.
We have given three concrete and loosely connected applications. First, we have 
given an independent proof that our operator algebra generates all zero modes of 
the class of Haldane pseudo-potential Hamiltonians whose highest filling factor 
ground states are the $1/M$-Laughlin states. Together with Ref. \onlinecite{Chen14},
this completes a program\cite{ortiz} that allows us to reproduces all known properties
of these zero-modes that are usually derived using analytic polynomial wave functions through
an alternative,
second quantized formalism.
The starting point here are the second quantized ``lattice'' versions of these pseudo-potential
Hamiltonians that are special instances of frustration free but {\em infinitely}-ranged
lattice models. We believe that our results will impact the more general development of 
such models which may have important applications in fractional Chern insulators.
\cite{Tang11,Sun11,Neu11,Sheng11,qi,Wang11,Regnault11,Ran11,Bernevig12,WangPRL12,WangPRB12,Wu12,Laeuchli12,Bergholtz12,Sarma12,Liu13,chen13, scaffidi}
In many ways our approach can be thought of as the {\em microscopic} bosonization of the 
lattice pseudo-potential Hamiltonians. Here, ``microscopic'' means
that the relation between boson modes creating exact zero energy eigenstates on the one hand
and the microscopic, second-quantized 
electron creation operator on the other is completely manifest. 
It may thus be no surprise that our approach also
leads to an alternative proof of one of the central results in bosonization, the equivalence of the
bosonic and the fermionic Fock spaces.
As a third application, we have extended some earlier results by Stone\cite{stone_book}
on the second quantized version of Read's  order parameter\cite{readOP}, or its ``string factor'' 
that creates local quasi-holes. We have given the explicit expression of this operator 
to the local electron creation operator and generalized its commutation relation with the latter
to the cylinder and spherical geometries.
We believe that the extension of our results to other models --both old and new-- will be 
an interesting direction for the future.

\begin{acknowledgments}
This work has been supported by the National Science
Foundation under NSF Grant No. DMR-1206781 (AS) and under NSF Grant No.
DMR-1106293 (ZN). AS would like to thank X. Tang for insightful
discussion. GO would like to acknowledge helpful discussions with M. Stone. 

\end{acknowledgments}

 \appendix
 
 \section{Newton-Girard operator relations between operators $\cO_d$ anD $e_d$}
\label{appA} 
 
In this Appendix we show explicitly that the operators $ e_{d} $ and $ \cO_{d} $ satisfy 
the {\it Newton-Girard operator} relations.
 Equations \eqref{Od} and \eqref{elementary2} are reproduced here for convenience as follows:

\be
   \cO_{d} = \sum _{r}  c_{r+d}^{\dagger}c_{r}^{\;}\,,
\ee

\be\label{edef}
      e_{d} = \frac{1}{d!} \sum _{ r_{1} \dotsc r_{d}}  c_{r_{1}+1}^{\dagger}c_{r_{2}+1}^{\dagger}\dotsm 
      c_{r_{d}+1}^{\dagger}c_{ r_{d}}^{\;}\dotsm c_{r_{2}}^{\;}c_{r_{1}}^{\;}\,,
\ee
The operators $e_d$ satisfy the commutation relations \eqref{ecomm} with electron
creation operators. Here we need the analogous relation for electron annihilation
operators:

\be\label{ecommA}
  \left[  e_{d}, c_{r}\right] = - e_{d-1} c_{r-1}\,.
\ee

From the definition \eqref{edef}, we also infer the following recursion relation,

\be
e_{d} = \frac{1}{d} \sum _{r}  c_{r+1}^{\dagger} e_{d-1}^{\;} c_{r}^{\;}\,,    
\ee
where again $e_0=1$.
Now using the commutator \eqref{ecommA} we can write this as:

\be
\begin{split}
e_{d} &= \frac{1}{d} \sum _{r}  c_{r+1}^{\dagger}  (c_{r}^{\;}e_{d-1}^{\;} - e_{d-2}^{\;} c_{r-1}^{\;})   
\\
& =  \frac{1}{d} (\cO_{1} e_{d-1} + (-1)\sum _{r}  c_{r+1}^{\dagger}   e_{d-2}^{\;} c_{r-1}^{\;})\,.
 \end{split}
 \ee
 By repeatedly applying the last step to the sum in the second line,
 we arrive at the following:
 
 \be
e_{d} = -\frac{1}{d} \sum _{ k\geq 1}  (-1)^k \cO_{k} e_{d-k} \,,
\ee
which is equivalent to the {\it Newton-Girard} relation \eqref{NG}.


\section{Proof of Eqs. \eqref{simplefacts}\label{dominanceApp}}

We consider \Eq{fact1} first, here reproduced as
\be\label{sf1}
\klambda\geq\klambdap \Leftrightarrow \ek_{\bar d} \klambda \geq \ek_{\bar d} \klambdap\,.
\ee
We identify $\klambda$, $\klambdap$ with partitions
$\{\lambda_i\}_{i=1\dotsc N}$, 
$\{\lambda'_i\}_{i=1\dotsc N}$, respectively.
Then $\ek_{\bar d} \klambda$, 
$\ek_{\bar d} \klambdap$, 
respectively, correspond to partitions 
$\{\ell_i\}_{i=1\dotsc N}$, 
$\{\ell'_i\}_{i=1\dotsc N}$,
 where
 \be
 \begin{split}
 \ell_i=\lambda_i+1\,,\;\;\ell_i'=\lambda'_i+1\quad &\mbox{for $i=1\dotsc d$,}\\
 \ell_i=\lambda_i\,,\;\;\ell_i'=\lambda'_i\quad &\mbox{for $i=d+1\dotsc N$.} 
 \end{split}
 \ee
 This clearly implies
 \be
 \sum_{i=1}^n (\ell_i-\ell'_i)= \sum_{i=1}^n (\lambda_i-\lambda'_i)\quad\mbox{for all $n=1\dotsc N$,}
 \ee
 and thus, by the criterion for dominance \eqref{domdef}, \Eq{sf1} follows.
 
 Now for \Eq{fact2}, reproduced here as
 \be\label{sf2}
   \ek_{\bar d} \klambda \geq \ek_S \klambda \quad\mbox{if $|S|=d$,}
\ee
we reiterate once more that, for fermions, we consider this equation automatically
satisfied if the right hand side vanishes (the left hand side can never vanish).
Let again $\klambda$, $\ek_{\bar d} \klambda$ be associated with partitions
 $\{\lambda_i\}_{i=1\dotsc N}$, , $\{\ell_i\}_{i=1\dotsc N}$, respectively, and let
 $\ek_{S} \klambda$ be associated with a partition  $\{\ell_i'\}_{i=1\dotsc N}$.
 Let 
\be\label{delta}
 \begin{split}
 \delta_i=1\quad &\mbox{for $i=1\dotsc d$,}\\
 \delta_i=0 \quad &\mbox{for $i=d+1\dotsc N$.} 
 \end{split}
 \ee
 Then 
 \be
     \ell_i=\lambda_i+\delta_i\,.
 \ee
 
Furthermore,  we can define numbers $s_i$, $i=1\dotsc N$ with
\be\label{characteristic}
    s_i=1\quad  \mbox {for $i\in S$,}\quad s_i=0\quad \mbox {for $i\notin S$},
\ee
i.e., $s_i$ is the characteristic function of the set $S$, such that
\be\label{ellp}
   \ell'_i=\lambda_i+s_i\,.
\ee
 We must convince ourselves that the $\ell'_i$ defined by the above equation may still
 be assumed to
 satisfy $\ell'_i\geq \ell'_{i+1}$, as long as the $\lambda_i$ satisfy the analogous relation.
 This can only become problematic for bosons, if $\lambda_i=\lambda_{i+1}$ and
 $s_i=0$, $s_{i+1}=1$. However, we may then instead consider a new sequence of numbers $s_i'$
 which is identical to $s_i$, except with the entries for $i$ and $i+1$ swapped.
 This is then the characteristic function of a set $S'$ different from $S$, where, however,
 $\ek_S$ and $\ek_{S'}$ have the same effect on $\klambda$. For this reason, we may assume
 without loss of generality that \Eq{ellp} leads to a monotonously decreasing set of numbers, as is must for the $\ell_i'$ to define a partition.
 Then the dominance criterion $\sum_{i=1}^n \ell_i \geq \sum_{i=1}^n \ell_i'$ is equivalent to
 \be
      \sum_{i=1}^n \delta_i \geq \sum_{i=1}^n s_i\,,
 \ee
 which, using the fact that $|S|=d$, evidently follows for all $n\leq N$ from Eqs. \eqref{delta}, \eqref{characteristic}.

 \bibliography{zero}
\end{document}